\documentstyle[twocolumn,pra,aps,epsfig]{revtex}
\tighten
\begin{document}
\preprint{}
\draft
\def\be{\begin{equation}}
\def\ee{\end{equation}}
\def\bq{\begin{eqnarray}}
\def\eq{\end{eqnarray}}

%
\title{Finite Temperature Quark Matter and Supernova Explosion}
\author{Alessandro Drago and Ubaldo Tambini}
\address{Dipartimento di Fisica, Universit\`a di Ferrara, 
and INFN, Sezione di Ferrara, Via Paradiso 12, Ferrara, Italy 44100\\}
\date{\today}
\maketitle
%
%
\begin{abstract}
We study the equation of state of quark matter at finite temperature,
using a confinement model in which 
chiral symmetry remains broken in the deconfined phase.
Implications for type II supernova explosion and for
the structure and evolution of the proto-neutron star are discussed.
\end{abstract}
%
%
%
%
\narrowtext

The Equation of State (EOS) of matter at high density and
temperature is a very interesting but controversial topic. Among the possible
applications of such an EOS are the structure of neutron stars (very
high density and low temperature), supernova explosions (high
density and moderate temperature) and relativistic heavy ions
collisions (high temperature and not very high density).

It is possible, at least in principle, to obtain information about
the behavior of matter under extreme conditions through lattice QCD
calculations.  Recently \cite{blum} it was shown that the
deconfinement transition seen at zero baryon density becomes a smooth
crossover at very small density and that at low enough temperature
chiral symmetry remains broken at all densities.  Although these
results need to be confirmed by other calculations, it is anyway
interesting to explore a model where these features are implemented.
In particular, if a constituent quark model is used, the deconfinement
transition need not to be as discontinuous as in MIT-like models,
where the transition to quark matter implies also chiral symmetry restoration.

In this letter we compute a finite temperature
EOS based on the transition to quark
matter, and study in particular its connection to the problem of
supernova explosion.  The most recent calculations show that using
traditional EOS a successful explosion is not
achieved {\it via} the prompt shock mechanism and late neutrino
transport still have to be more exhaustively investigated
\cite{arnett}. The prompt explosion mechanism
needs a very soft EOS, what seems apparently incompatible with the observed
masses of neutron stars.  
The last conclusion is based essentially on the results coming from the
phenomenological BCK EOS \cite{bck,brown}.
This difficulty can be circumvented.
In order to obtain a
successful explosion one needs a softening of the EOS at densities
slightly larger than the nuclear matter saturation density
$\rho_0$\cite{swesty}. On the other hand,
the maximum mass of a neutron star depends 
on the very large density behavior of the EOS and requires a not too
soft EOS in that density region.

The idea to correlate the softening of the EOS to the presence of
a phase transition\cite{migdal} and in particular to the formation
of quark matter\cite{takahara} is rather old. 
This possibility,
in connection with the problem of supernova explosion,
has been considered recently by Gentile {\it et
al.}\cite{gentile} in a 
phenomenological way, without any attempt to relate the parameters
governing the transition to other quark model calculations. 
Moreover the dependence of the EOS on the temperature and on the
electron fraction was not investigated. 
Since during the first seconds of the its life a proto-neutron
star deleptonizes, it is particularly important to study how 
the transition to quark matter is affected by the electron fraction.

We present here a calculation of the transition to quark matter at finite
temperature. In order to satisfy Gibbs conditions when more than one
charge is conserved, one has to use the technique developed by
Glendenning \cite{glenpt}.  Until now, the only calculation
incorporating this technique   at finite temperature was the study of the
liquid-gas phase transition in nuclear matter made by M\"uller and
Serot \cite{serot}. 

We will use the Color Dielectric Model (CDM)
\cite{pirner92,birse90,su3col}
to describe quark matter.  The CDM is a confinement model which has
been used with success to study properties of single nucleons, such as
structure functions \cite{barone} and form factors \cite{ff},
or to investigate zero temperature 
quark matter \cite{rosina} and the structure of
neutron stars \cite{jensen}.

An important feature of the CDM is that effective quark masses are
of the order of 100 MeV even at very large
densities, hence chiral
symmetry is broken and the Goldstone bosons are relevant degrees of
freedom. This is to be contrasted with models like the MIT bag, where
quarks have masses of a few MeV.

The Lagrangian of the model reads:
\begin{eqnarray}
     {\cal L} &=& i\bar \psi \gamma^{\mu}\partial_{\mu} \psi 
 +{1\over 2}{\left(\partial_\mu\sigma \right)}^2
     +{1\over 2}{\left(\partial_\mu\vec\pi\right)}^2
     -U\left(\sigma ,\vec\pi\right)   \nonumber\\
     &+&\!\!\!\sum_{f=u,d} {g_f\over f_\pi \chi} \, \bar \psi_f\left(\sigma
     +i\gamma_5\vec\tau\cdot\vec\pi\right) \psi_f 
     +{g_s \over \chi} \, \bar \psi_s \psi_s       \nonumber
     \\
      &+&{1\over 2}{\left(\partial_\mu\chi\right)}^2
      -{1\over 2}{\cal M}^2\chi^2
\end{eqnarray}
where $U(\sigma ,\vec\pi)$ is the ``mexican-hat'' potential, as in
Ref.\ \cite{ff}.
The Lagrangian ${\cal L}$  
describes a system of interacting $u$, $d$ and $s$ quarks, pions, sigmas and
a scalar--isoscalar chiral singlet field $\chi$.
The scalar field $\chi$ characterizing the CDM is related
to the fluctuation of the gluon condensate around its 
vacuum expectation value\cite{pirner92}.
In the model this fluctuation is rather small indicating a smooth 
transition between the exterior and the interior of the nucleon and 
allowing a soft deconfinement.

The coupling constants are given by $g_{u,d}=g (f_{\pi}\pm \xi_3)$
and $g_s= g(2 f_K -f_{\pi})$, where $f_{\pi}=93$ MeV and $f_{K}=113$ MeV 
are the pion and the kaon 
decay constants, respectively, and $\xi_3=f_{K^\pm}-f_{K^0}=-0.75$ MeV. 
These coupling constants depend on a single parameter $g$.

Confinement is obtained {\it via} the effective quark masses
$m_{u,d}=g_{u,d} \bar\sigma/(\bar\chi f_\pi)$ and 
$m_s=g_s / \bar\chi$,
which diverge outside the nucleon.
Indeed, the classical fields
$\bar \chi$ and $\bar\sigma$ are solutions of the Euler--Lagrange equations
and
$\bar\chi$ goes asymptotically to zero at large distances.

In the following we will use for the model parameters the values: 
\be
g=0.023\, {\rm GeV}\,, \quad {\cal M}=1.7\, {\rm GeV} \,,
\ee
giving a nucleon isoscalar radius of 0.80 fm 
(exp.val.=0.79 fm) and an average
delta--nucleon mass of 1.129 GeV (exp.val.=1.085 GeV). A similar set
of parameters has been used to compute structure functions \cite{barone}
and form factors \cite{ff} and to study neutron stars \cite{jensen}.

We describe the hadronic phase with a relativistic field
theoretic model of the Walecka type \cite{sw86}, 
including protons and neutrons
only. The parameters used to define the Lagrangian of the hadronic part
are the ones labeled HS81
in the work by Knorren {\em et al.} \cite{kpe95}.

The transition to quark matter has been studied using the technique
developed by Glendenning\cite{glenpt}, since in the transition
two quantities are conserved, the baryon
(B) and the electric (C) charge :
\bq 
\rho_B&=&(1-\chi)\rho_B^h+\chi\rho_B^q\\
\rho_C&=&(1-\chi)\rho_C^h+\chi\rho_C^q+\rho_e+\rho_\mu=0\,\nonumber .
\eq
Here $\chi$ is the fraction of matter in the quark phase. The 
superscripts $h$ and $q$ label the density in the hadronic and in the
quark phase, respectively. The electron ($\rho_e$) and the muon
($\rho_\mu$) charge densities 
contribute to make the total electric charge equal to zero.
Due to the presence of strange quarks electron and muon densities are
suppressed.

Since matter has to be in chemical equilibrium under $\beta$-decay and
deconfinement, the following equations have to be satisfied:
\bq
\mu_n-\mu_p=\mu_e-\mu_{\nu_e}\,&,&\quad
\mu_n-\mu_p=\mu_\mu-\mu_{\nu_\mu}\nonumber\\
2\mu_d+\mu_u=\mu_n\,&,&\quad\mu_u-\mu_d=\mu_p-\mu_n\nonumber\\
\mu_s&=&\mu_d\,,
\eq
together with the usual condition for mechanical equilibrium, {\it i.e.}
the equality of the pressure in the two phases:
\be
P^h=P^q \,.
\ee

Other conditions depend on the specific problem under discussion.
For instance, in order
to study the structure of the star after deleptonization, one assumes
that neutrinos can escape freely, so
their chemical potential is set to zero
in previous equations. This assumption is incorrect 
in the first seconds of the life
of the proto-neutron star. Due to neutrino opacity 
lepton numbers are conserved, neutrino chemical potentials
are different from zero and the EOS is computed for
fixed values of the lepton fractions:
\be
Y_{l_e}=(\rho_e+\rho_{\nu_e})/\rho_B\,,\quad
Y_{l_\mu}=(\rho_\mu+\rho_{\nu_\mu})/\rho_B\,.
\ee
Since before the collapse $Y_{l_\mu}=0$, this
quantity has been kept fixed and
the EOS has been computed for various values of $Y_{l_e}$.

In the computation of the EOS
all previous equations have been solved together with the mean-field equations
of the Walecka model and of the CDM. Finite temperature has been taken into
account using the standard technique developed
{\it e.g.} in Ref.\cite{kapusta}.

\begin{figure}
\centerline{\hbox{
\psfig{figure=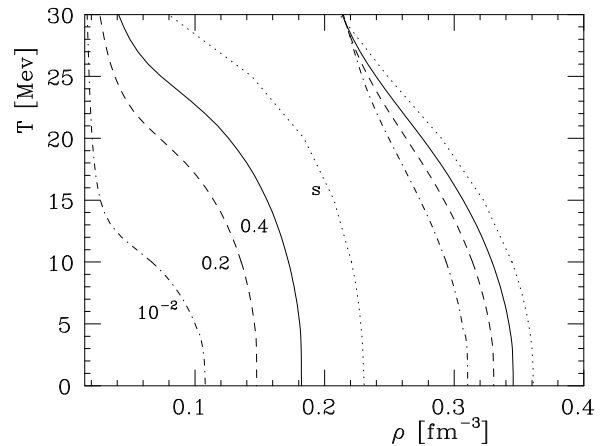,angle=90,width=8cm}
}}
\caption{Boundaries separating hadronic matter, mixed phase and quark matter
in the density-temperature plane.
The labels indicate various values of $Y_{l_e}$, $s$ is for symmetric matter.
}
\label{fig:tran}
\end{figure}

We come now to the results.
In Fig.\ref{fig:tran} we present
the boundaries separating hadronic matter from
mixed phase and the latter from pure quark matter.
The labels correspond to various values of $Y_{l_e}$.
Symmetric nuclear matter is also presented.
The transition region depends on the electron fraction $Y_{l_e}$.
In symmetric matter at low temperatures the mixed phase forms at
$\rho=0.23$ fm$^{-3}$, 
therefore no quark matter is present in heavy nuclei.
Decreasing the value of $Y_{l_e}$  
the phase transition starts at lower densities.
At any value of $Y_{l_e}$
the mixed phase extends on a rather limited range of densities
and even at zero temperature
pure quark matter phase is reached 
at densities slightly larger than $2 \rho_0$.
At higher temperatures the transition starts at lower densities. We have
explored only the region of moderate temperatures, the relevant one
for supernova explosion. 
In order to study higher temperatures it would have been
essential to take into account quantum corrections beyond 
mean field approximation.

The {\it scenario} depicted in Fig.\ref{fig:tran} is rather 
different from the one described by the MIT model, where the transition,
at least in symmetric nuclear matter, 
starts at very large densities and the mixed phase
extends on a broad density range. Using the MIT model, 
which assumes the interior of the nucleon to be in a perturbative regime,
it is almost unavoidable to have a first order transition.
In the CDM it is conceivable, at least in principle, 
to obtain a smoother
transition by taking into account
quark correlations beyond mean field approximation. 

\begin{figure}
\centerline{\hbox{
\psfig{figure=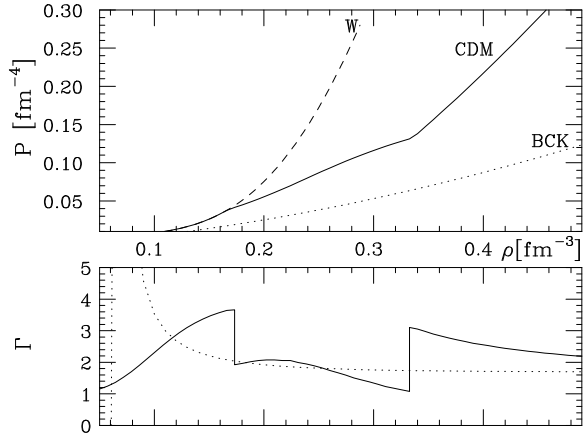,angle=90,width=8cm}
}}
\caption{Pressure (upper box) and adiabatic index (lower box) as function
of the density in CDM (solid) and in BCK (dotted). The pressure in Walecka
model is also shown (dashed).
}
\label{fig:eos}
\end{figure}
    
To investigate our EOS in connection with the problem of supernova
explosion, we compare with BCK EOS\cite{bck} (parameters as in model
38). The latter is a totally phenomenological EOS which is soft enough
to allow for supernova explosion, but gives a maximum mass smaller
than the mass of PSR 1913+16 (1.44 $M_\odot$)\cite{brown}.  In our
model the maximum value of the gravitational mass for a non-rotating
cold star is 1.59 $M_\odot$\cite{jensen}.  

In Fig.\ref{fig:eos} we
present results for $Y_{l_e}=0.4$ and entropy per baryon number
$S/R=1$.  Due to the presence of strange quarks, the electron fraction
$Y_e$ is rather small in the quark phase, $Y_e\sim0.3$, and the muon
fraction $Y_\mu$ is always very small.  In the upper box we
compare the pressure in the Walecka model, in our model and in BCK
EOS.  Due to the phase transition, our EOS is rather soft from $\rho=0.17$
fm$^{-3}$ to $\rho=0.34$ fm$^{-3}$.  On the other hand, after $\rho =
0.34$ fm$^{-3}$ it is considerably stiffer than BCK, allowing higher masses
for the proto-neutron star.  These conclusions are strengthened by the
computation of the adiabatic index, shown in the lower box of 
Fig.\ref{fig:eos}.
Clearly in the mixed phase matter offers little resistance to
collapse, but when pure quark matter phase is reached the collapse is
halted. In the mixed phase region our adiabatic index is even smaller
than in BCK.

Another way of investigating the possibility of a prompt explosion using
our EOS is to estimate the
\begin{table}
\begin{tabular}{cccc}
           &  $BE$ &       $V_g$   &    $ E_i$ \\
\tableline
Walecka   &   24 &       -73   &     97 \\

CDM       &   32  &      -93     &   125 
\end{tabular}
\caption{Total binding energy, gravitational and internal energies
in Walecka model and in CDM (see text). All energies are in foe
($10^{51}$erg).} 
\end{table}
\noindent extra energy at disposal for the shock wave in our
model respect to pure Walecka model.
To this purpose we follow the analysis of
Gentile {\it et al.}\cite{gentile}.
Their way of estimating the extra energy is the following:
they compute the binding energy of the central part of the proto-neutron
star, immediately before bounce, integrating on a mass of $0.5 M_\odot$.
The binding energy is the sum of gravitational and internal potential energy:
$BE = V_g + E_i$ = gravitational mass -- baryonic mass.
Integrating only on the inner region,
a positive value for the $BE$ is obtained, indicating that this region tries
to expand, pushing the exterior envelope. The larger the (anti-)binding
energy, the stronger the push. We can therefore compare these numbers
in Walecka and in our model, using the EOS with $S/R=1$ and $Y_{l_e}=0.4$.
The result is shown in the Table. An extra energy of about
8 foe is at disposal for the explosion.

\begin{figure}
\centerline{\hbox{
\psfig{figure=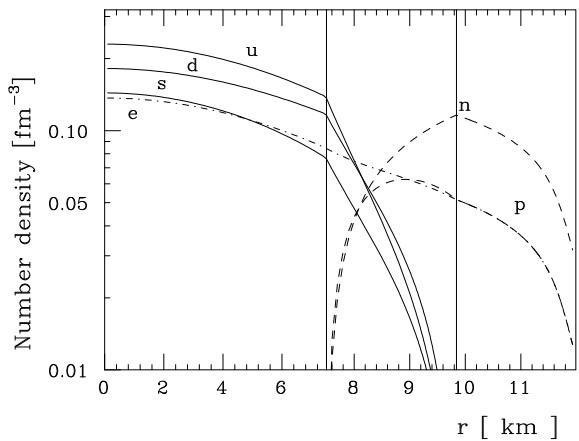,angle=90,width=8cm}
}}
\caption{Composition of a star having $S/R=1$ and $Y_{l_e}=0.4$. 
A different length's scale has been used where the mixed phase is formed.
}
\label{fig:compo}
\end{figure}

\begin{figure}
\centerline{\hbox{
\psfig{figure=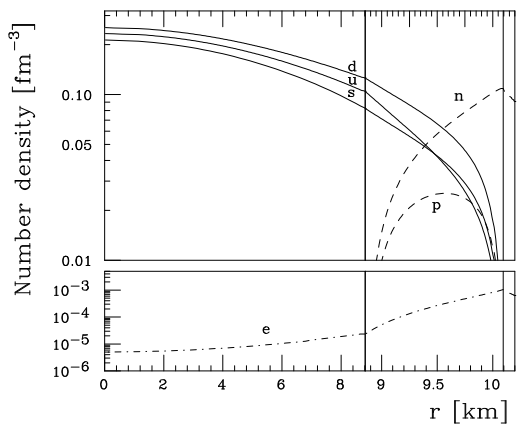,angle=90,width=8cm}
}}
\caption{Same as Fig.\ref{fig:compo}, for a star having $S=0$ and
$\beta$-stable.
}
\label{fig:compo2}
\end{figure}

We come now to the second problem we like to study, namely the
structure and evolution of the proto-neutron star.
In Fig.\ref{fig:compo} we show the composition of a star
with  $Y_{l_e} = 0.4$ 
($Y_e \sim 0.3$) and entropy {\it per} particle $S/R = 1$.
The central temperature is of the order of 10 MeV.
These conditions should be realized at the bounce.
Immediately after the bounce the entropy increases. After a time of the 
order of 10 seconds the proto-neutron star cools down and deleptonizes.
After that the composition of the star does not change any more. We show this
later stage in Fig.\ref{fig:compo2}, where we assume entropy $S = 0$
and $\beta$-stability. 
In both figures the baryonic mass is
the same, $1.54 M_\odot$, which corresponds to a gravitational mass of
$1.4 M_\odot$ for the final star of Fig.\ref{fig:compo2}.

\begin{figure}
\centerline{\hbox{
\psfig{figure=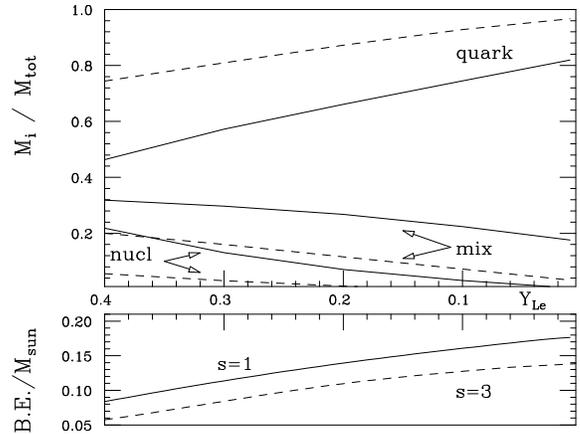,angle=90,width=8cm}
}}
\caption{Mass fractions in hadronic, mixed and quark phase (upper box)
and binding energy (lower box) as function of $Y_{l_e}$ for 
entropy $S/R=1$ (solid) and $S/R=3$ (dashed), respectively.
}
\label{fig:frac}
\end{figure}

There have been many speculations about the possibility of a late
neutrino emission, based on the SN1987 data \cite{kahana}.  In
particular the idea of a late transition to strange matter with a new
emission of energetic neutrinos has been invoked. We investigate this
possibility in Fig.\ref{fig:frac}, where we show the composition of
the star and its binding energy as a function of the lepton fraction
$Y_{l_e}$ and for two values of the entropy {\it per} particle. To
study this problem one should really solve the dynamics' equations
giving lepton fraction and entropy as a function of time. What can be
learnt from Fig.\ref{fig:frac} is the absence of a sudden jump in the
composition of the star and in its binding energy as a function of
$Y_{l_e}$.  The conclusion we can draw is that an emission of
energetic neutrinos during the first seconds is indeed possible, since
the binding energy increases steadily as $Y_{l_e}$ is
diminishing. Anyway a peak in the neutrino luminosity can be due only
to the dynamics of the explosion, not to a discontinuity in the
EOS of the proto-neutron star.

Recently there has been a speculation about the possibility of having
a proto-neutron star collapsing later to a black hole,
due to the softening of the EOS after deleptonization\cite{lat}. 
In our calculation we do not have
this effect, because the maximum baryonic mass for the proto-neutron
star is essentially independent of the value of $Y_e$ and $S$ and
remains always near $1.82 M_\odot$.

In this letter we have shown the possibility to have a phase
transition at densities slightly larger than $\rho_0$ in pre-supernova
matter. This transition softens the EOS which in this range of
densities is comparable with the phenomenological BCK EOS, but gives
acceptable values for the maximum gravitational mass of the final
neutron star. Our EOS indicates also the possibility of a energetic
neutrino emission in the first seconds of the life of the
proto-neutron star. All these conclusions will soon be tested in a
dynamical simulation of the explosion.

It is a pleasure to thank V.Barone, Z.Berezhiani, L.Caneschi, H.-T.Janka and
W.Keil for many stimulating discussions.

%

%


\begin{references}

\bibitem{blum} T.\ Blum, J.E.\ Hetrick and D.\ Toussaint, 
Phys.\ Rev.\ Lett.\ {\bf 76}, 1019 (1996).
\bibitem{arnett}D.Arnett, {\it Supernovae and Nucleosynthesis},
Princeton University Press, Princeton 1996.
\bibitem{bck}E.A.Baron, J.Cooperstein and S.Kahana, Phys. Rev. Lett. {\bf 55},
126 (1985).
\bibitem{brown}G.E.\ Brown, Phys. Rep. {\bf 163}, 167 (1988).
\bibitem{swesty}F.D.Swesty, J.M.Lattimer and E.S.Myra, Astrophys. J.
{\bf 425}, 195 (1994).
\bibitem{migdal}A.B.Migdal, A.I.Cherenoutsan and I.N.Mishustin,
Phys. Lett. {\bf B83}, 158 (1979).
\bibitem{takahara}M.Takahara and K.Sato, Phys. Lett. {\bf B156}, 17 (1985).
\bibitem{gentile}N.A.\ Gentile, \ M.B.\ Aufderheide, \ G.J.\ Mathews, 
\ F.D.\ Swesty
and G.M.Fuller, Astrophys. J. {\bf 414}, 701 (1993).
\bibitem{glenpt}N.K.\ Glendenning, \ Phys.\ Rev.\  {\bf D46}, 1274 (1992).
\bibitem{serot}H.\ M\"uller and B.D.\ Serot,\ Phys.\ Rev.\ {\bf C52}, 
2072 (1995).
\bibitem{pirner92} H.J.\ Pirner, Prog.\ Part.\ Nucl.\ Phys.\  {\bf 29},
33, (1992).
\bibitem{birse90} M.C.\ Birse, Prog.\ Part.\ Nucl.\ Phys.\  {\bf 25},
1 (1990).
\bibitem{su3col}J.A.\ McGovern,\ Nucl.\ Phys.\  {\bf A533}, 553 (1991).
\bibitem{barone}V.\ Barone\ and A.\ Drago,\ Nucl.\ Phys.\  {\bf A552},
479 (1993);  {\bf A560}, 1076 (1993);
V.\ Barone,\ A.\ Drago and M.\ Fiolhais,\ Phys.\ Lett.\ {\bf B338}, 433 (1994).
\bibitem{ff}A.\ Drago, M.\ Fiolhais,\ U.\ Tambini, Nucl. Phys.
 {\bf A609}, 488 (1996).
\bibitem{rosina}W.\ Broniowski,\ M.\ \^Cibej,\ M.\ Kutschera\ and M.\
Rosina,\ Phys.\ Rev.\ {\bf D41} (1990) 285; A.\ Drago, A.\ Fiolhais and U.\ Tambini,
Nucl.\ Phys.\  {\bf A588}, 801 (1995).
\bibitem{jensen}A. Drago, U. Tambini, M.Hjorth-Jensen, Phys. Lett.
{\bf B380}, 13 (1996).
\bibitem{sw86} B.D.Serot and J.D.Walecka, Adv.Nucl.Phys.{\bf 16}(1986)1.
\bibitem{kpe95} R.\ Knorren, M.\ Prakash and P.J.\ Ellis,
Phys.\ Rev.\ {\bf C52}, 3470 (1995).
\bibitem{kapusta} J.I.\ Kapusta, 
{\it Finite-temperature field theory}, 
Cambridge University Press, Cambridge 1989.
\bibitem{kahana} S.H. Kahana, Ann. Rev. Nuc. Part. Sci. {\bf 39}, 231 (1989).
\bibitem{lat}M.\ Prakash, I.\ Bombaci, M.\ Prakash,
P.J.\ Ellis, J.M.\ Lattimer and R.\ Knorren, nucl-th/9603042,
to appear in Phys.Rep.

\end{references}
\end{document}